\documentclass[draftcls, onecolumn, 11pt]{IEEEtran} % [draft] draftcls, onecolumn, 12pt or [journal [draftcls, onecolumn, 12pt]
\usepackage {graphicx}  % Written by David Carlisle and Sebastian Rahtz
\usepackage {amssymb}
\usepackage {amsmath}
\usepackage {mathrsfs}

\usepackage{amsmath}   % From the American Mathematical Society
\begin{document}
%\title{Linear Programming Equalization}
\title{Graph-Based Decoding in the Presence of ISI}
\author{\authorblockN{Mohammad H. Taghavi and Paul H. Siegel}\\
\authorblockA{Center for Magnetic Recording Research\\
University of California, San Diego\\
La Jolla, CA 92093-0401, USA\\
Email: (mtaghavi, psiegel)@ucsd.edu}
}% <-this % stops a space}
%
%

\maketitle

\begin{abstract}
We propose an approximation of maximum-likelihood detection in ISI channels based on linear programming or message passing. We convert the detection problem into a binary decoding problem, which can be easily combined with LDPC decoding. We show that, for a certain class of channels and in the absence of coding, the proposed technique provides the exact ML solution without an exponential complexity in the size of channel memory, while for some other channels, this method has a non-diminishing probability of failure as SNR increases. Some analysis is provided for the error events of the proposed technique under linear programming.
\end{abstract}

% \begin{keywords}
% ISI channels, maximum-likelihood detection, linear programming, iterative message passing, combine equalization and decoding.
% \end{keywords}

% Will be ignored for other modes.
% \IEEEpeerreviewmaketitle

\newtheorem{definition}{Definition}
\newtheorem{notation}{Notation}
\newtheorem{theorem}{Theorem}
\newtheorem{lemma}{Lemma}
\newtheorem{claim}{Claim}
\newtheorem{algorithm}{Algorithm}
\newtheorem{corollary}{Corollary}
\newtheorem{conjecture}{Conjecture}
\newtheorem{remark}{Remark}

\section{Introduction}
Intersymbol interference (ISI) is a characteristic of many data communications and storage channels. Systems operating on these channels employ error-correcting codes in conjunction with some ISI reduction technique, which, in magnetic recording systems, is often a conventional Viterbi detector. It is known that some gain will be obtained if the equalization and decoding blocks are combined at the receiver by exchanging soft information between them. A possible approach to achieving this gain is to use soft-output equalization methods such as the BCJR algorithm \cite{BCJR} or the soft-output Viterbi algorithm (SOVA) \cite{SOVA} along with iterative decoders. However, both BCJR and SOVA suffer from exponential complexity in the length of the channel memory.

%Kurkoski \emph{et al.} \cite{Kurkoski} proposed a bit-based and a state-based graph representation of the ISI channel that can be combined with the Tanner graph of a low-density parity-check (LDPC) code for joint message-passing (MP) decoding. They showed that the bit-based method suffers from a significant performance degradation due to the abundance of 4-cycles, but the state-based method has a performance and overall complexity similar to BCJR, while benefiting from a parallel structure and reduced delay.
Kurkoski \emph{et al.} \cite{Kurkoski} proposed two graph representations of the ISI channel that can be combined with the Tanner graph of the LDPC code for message-passing decoding. Their bit-based representation of the channel contains many 4-cycles, which results in a significant performance degradation compared to maximum-likelihood (ML) detection. On the other hand, message passing (MP) on their state-based representation, where messages contain state rather than bit information, has a performance and overall complexity similar to BCJR, while benefiting from a parallel structure and reduced delay. 
Among other works, Singla \emph{et al.} \cite{Singla} applied message passing on a bit-based graph representation of a two-dimensional ISI channel combined with an LDPC Tanner graph. However, similar to the case of one-dimensional ISI, the abundance of short cycles prevents the algorithm from performing close to optimal.

Linear programming (LP) has been recently applied by Feldman \emph{et al.} \cite{Feldman} to the problem of ML decoding of LDPC codes, as an alternative to MP techniques. In this method, the binary parity-check constraints of the code are relaxed to a set of linear constraints in the real domain, thus turning the integer problem into an LP problem. While LP decoding performs closely to MP algorithms such as the sum-product algorithm (SPA) and the min-sum algorithm (MSA), it is much easier to analyze for finite code lengths.

Motivated by the success of LP decoding, in this work we study the problem of ML detection in the presence of ISI, which can be written as an integer quadratic program (IQP). We convert this problem into a binary decoding problem, which can be used for MP decoding, or, after relaxing the binary constraints, LP decoding. Furthermore, decoding an underlying LDPC code can be incorporated into this problem simply by including the parity checks of the code. 
%This type of LP relaxation has been used in various forms in integer programming applications \cite{Watters}.

By a geometric analysis we show that, in the absence of coding, if the impulse response of the ISI channel satisfies certain conditions, the proposed LP relaxation is guaranteed to produce the ML solution at all SNR values. This means that there are ISI channels, which we call \emph{LP-proper} channels, for which uncoded ML detection can be achieved with a complexity polynomial in the channel memory size. On the other end of the spectrum, some channels are \emph{LP-improper}, i.e. the LP method results in a nonintegral solution with a probability bounded away from zero at all SNR, even in the absence of noise. Furthermore, we observe some intermediate \emph{asymptotically LP-proper} channels where the performance asymptotically converges to that of ML detection at high SNR. When message passing is used instead of LP, we observe a similar behavior. Moreover, when LDPC decoding is incorporated in the detector, LP-proper channels achieve very good performance, while some other channels cannot go below a certain word error rate (WER).

The rest of this paper is organized as follows. In Section II, we describe the channel, and introduce the LP relaxation of ML detection. The performance analysis and simulation results of uncoded graph-based detection are presented in Section III. In Section IV, we study the combination of graph-based detection and LDPC decoding, and Section V concludes the paper.

\section{Relaxation of the Equalization Problem}
\subsection{Channel Model}

\begin{figure}
\centering
\includegraphics[width=4.0 in] {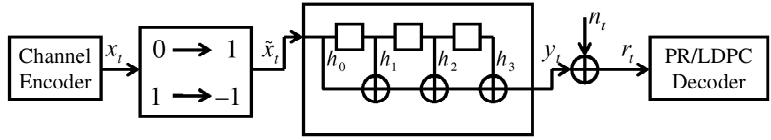}
\caption{Binary-input ISI channel.}\!\!\!\!\!\!\!\!\!
\label{ISI channel}
\end{figure}

We consider a partial-response (PR) channel with bipolar (BPSK) inputs, as described in Fig. \ref{ISI channel}, and use the following notation for the transmitted symbols.
\begin{notation}
The bipolar version of a binary symbol, $b\in\{0,1\}$, is denoted by $\tilde{b} \in \{-1,1\}$, and is given by
\begin{equation}
\label{bipolar}
\tilde{b} = 1-2b.
\end{equation}
\end{notation}

The partial-response channel transfer polynomial is $h(D)=\sum_{i=0}^{\mu} h_i D^i$, where $\mu$ is the channel memory size. Thus, the output sequence of the PR channel in Fig. \ref{ISI channel} before adding the white Gaussian noise can be written as
\begin{equation}
\label{y_t}
y_t = \sum_{i=0}^{\mu} h_i \tilde{x}_{t-i}.
\end{equation}

\subsection{Maximum-likelihood (ML) Detection}
%The objective of the ML detector is to find the codeword which minimizes the Euclidean distance between vector $\underline{y}=[y_1\ y_2\ \cdots \ y_n]^T$ and the vector of received samples, $\underline{r}$. In other words, it solves the optimization problem
Having the vector of received samples $\underline{r}=[r_1\ r_2\ \cdots \ r_n]^T$, the ML detector solves the optimization problem
\begin{align}
\label{MLD}
&\text{Minimize} \hspace{0.3 in} \left\| \underline{r}-\underline{y} \right\|_2 \nonumber \\
&\text{Subject to} \hspace{0.3 in} \underline{x} \in \mathscr{C},
\end{align} 
where $\mathscr{C} \subset\{0,1\}^n$ is the codebook and $\|\cdot\|_2$ denotes the $L_2$-norm. By expanding the square of the objective function, the problem becomes equivalent to minimizing
\begin{align}
\label{expand MLD}
\sum_t (r_t - y_t)^2 = & \sum_t \left[ r_t^2 - 2r_t\sum_i h_i \tilde{x}_{t-i} + \left(\sum_i h_i \tilde{x}_{t-i} \right)^2 \right] \nonumber\\ 
= & \sum_t \bigg[ r_t^2 - 2r_t\sum_i h_i \tilde{x}_{t-i} + \sum_i h_i^2 \tilde{x}_{t-i}^2 \nonumber\\
& + \mathop{\sum\sum}\limits_{i\neq j} h_i h_j \tilde{x}_{t-i}\tilde{x}_{t-j} \bigg],
\end{align}
where, for simplicity, we have dropped the limits of the summations. Equivalently, we can write the problem in a general matrix form
\begin{align}
\label{MLD matrix form}
&\text{Minimize} \hspace{0.3 in} -\underline{q}^T \underline{\tilde{x}} + \frac{1}{2} \underline{\tilde{x}}^T P \underline{\tilde{x}}, \nonumber\\
&\text{Subject to} \hspace{0.3 in} \underline{x} \in \mathscr{C},
\end{align} 
where in this problem $q_t = \sum_i h_i r_{t+i}$, and $P=H^TH$, with $H$ defined as the $n \times n$ Toeplitz matrix
\begin{equation}
\label{Toeplitz}
H=
\begin{bmatrix}
h_0 & 0 &\cdots\\
\vdots & \ddots\\
h_\mu & \cdots & h_0 & 0 \\
0   & h_\mu & & h_0 & 0\\
\vdots & & \ddots & & \ddots\\
0 & \cdots & 0 & h_\mu & \cdots &  h_0
\end{bmatrix}.
\end{equation}
Here we have assumed that $\mu$ zeros are padded at the beginning and the end of the transmitted sequence, so that the trellis diagram corresponding to the ISI channel starts and ends at the zero state.
If the signals are uncoded, i.e. $\mathscr{C}=\{0,1\}^n$, and $\underline{q}$ and $P$ are chosen arbitrarily, (\ref{MLD matrix form}) will represent the general form of an integer quadratic programming (IQP) problem, which is, in general, NP-hard. In the specific case of a PR channel, where we have the Toeplitz structure of (\ref{Toeplitz}), the problem can be solved by the Viterbi algorithm with a complexity linear in $n$, but exponential in $\mu$. However, this model can also be used to describe other problems such as detection in MIMO or two-dimensional ISI channels. Also, when the source symbols have a non-binary alphabet with a regular lattice structure such as the QAM and PAM alphabets, the problem can be reduced to the binary problem of (\ref{MLD matrix form}) by introducing some new variables.

\subsection{Problem Relaxation}
A common approach for solving the IQP problem is to first convert it to an integer LP problem by introducing a new variable for each quadratic term, and then relax the integrality condition; e.g. see \cite{Watters}. While this relaxed problem does not necessarily have an integer solution, it can be used along with branch-and-cut techniques to solve integer problems of reasonable size.
A more recent method is based on dualizing the IQP problem twice to obtain a convex relaxation in the form of a semi-definite program (SDP) \cite{Lemarechal}\cite{Goemans}. 
%A more recent method to deal with the IQP problem takes advantage of the fact that the Lagrangian dual of any optimization problem is a convex problem. Hence, if we dualize the integrality constraints in the IQP, and then dualize the result again, we will obtain a relaxed and convex version of the original problem, which will be in the form of a semi-definite program (SDP) \cite{Lemarechal}\cite{Goemans}. Random projection methods have been shown to provide effective ways to move back to the original feasible set, once a non-integral relaxed solution is found \cite{Goemans}.

In this work, we use the linear relaxation due to the lower complexity of solving LPs compared to SDPs. Unlike in \cite{Watters}, where the auxiliary variables are each defined as the product of two 0-1 variables, we define them as the product of $\pm 1$ variables, which, as we will see, translates into the modulo-2 addition of two bits when we move to the 0-1 domain. This relaxation is more suitable for our purpose, since modulo-2 additive constraints are similar to parity-check constraints; thus, message-passing decoders designed for linear codes can be applied without any modification. However, it can be shown that this relaxation gives the exact same solution as in \cite{Watters}.

To linearize (\ref{expand MLD}), we define
\begin{equation}
\tilde{z}_{t,j}=\tilde{x}_t \cdot \tilde{x}_{t-j},\ \ j=1,\ldots,\mu,\ t=j+1,\ldots,n.
\end{equation}
In the binary domain, this will be equivalent to
\begin{equation}
z_{t,j}=x_t \oplus x_{t-j},
\end{equation}
where $\oplus$ stands for modulo-2 addition. Hence, the right-hand side of (\ref{expand MLD}) is a linear combination of $\{x_t\}$ and $\{z_{t,j}\}$, plus a constant, given that $\tilde{x_i}^2 = 1$ is a constant. With some simplifications, the IQP in (\ref{MLD matrix form}) can be rewritten as
\begin{align}
\label{ILP}
\text{Minimize} \hspace{0.13 in} &\sum_t q_t x_t + \sum_t \sum_j \lambda_{t,j} z_{t,j}, \nonumber\\
\text{Subject to} \hspace{0.13 in} &\underline{x} \in \mathscr{C}, \nonumber\\
&z_{t,j}=x_t \oplus x_{t-j},\ j=1,\ldots,\mu, \nonumber\\
&\hspace{1.07 in} t=j+1,\ldots,n,
\end{align}
where, in the equalization problem,
\begin{equation}
\label{lambda}
\lambda_{t,j}=-P_{t,t-j}=-\!\!\! \sum_{i=0}^{\min(\mu-j, n-t)} h_i h_{i+j}.
\end{equation}
In this optimization problem, we call $\{x_i\}$ the information bits, and $\{z_{t,j}\}$ the state bits. It can be seen from (\ref{lambda}) that $\lambda_{t,j}$ is independent of $t$, except for indices near the two ends of the block; i.e.  $1\leq t \leq \mu$ and $n-\mu+1 \leq t \leq n$. In practice, this ``edge effect'' can be neglected due to the zero padding at the transmitter. For clarity, we sometimes drop the first subscript in $\lambda_{t,j}$, when the analysis is specific to the PR detection problem.
%In this optimization problem, we call $\{x_i\}$ the information bits, and $\{z_{t,j}\}$ the state bits. It can be seen from (\ref{lambda}) that $\lambda_{t,j}$ is independent of $t$, except for indices near the two ends of the code; i.e.  $1\leq t \leq \mu$ and $n-\mu+1 \leq t \leq n$. In practice, this ``edge effect'' can be neglected, since $\mu$ zeros are padded to the beginning and the end of the sequence, in order to make sure the trellis diagram of the PR channel starts from and ends in the zero state. Hence, for clarity, we sometimes drop the first subscript in $\lambda_{t,j}$, when the analysis is specific to the PR detection problem.

\begin{figure}
\centering
\includegraphics[width=4.0 in] {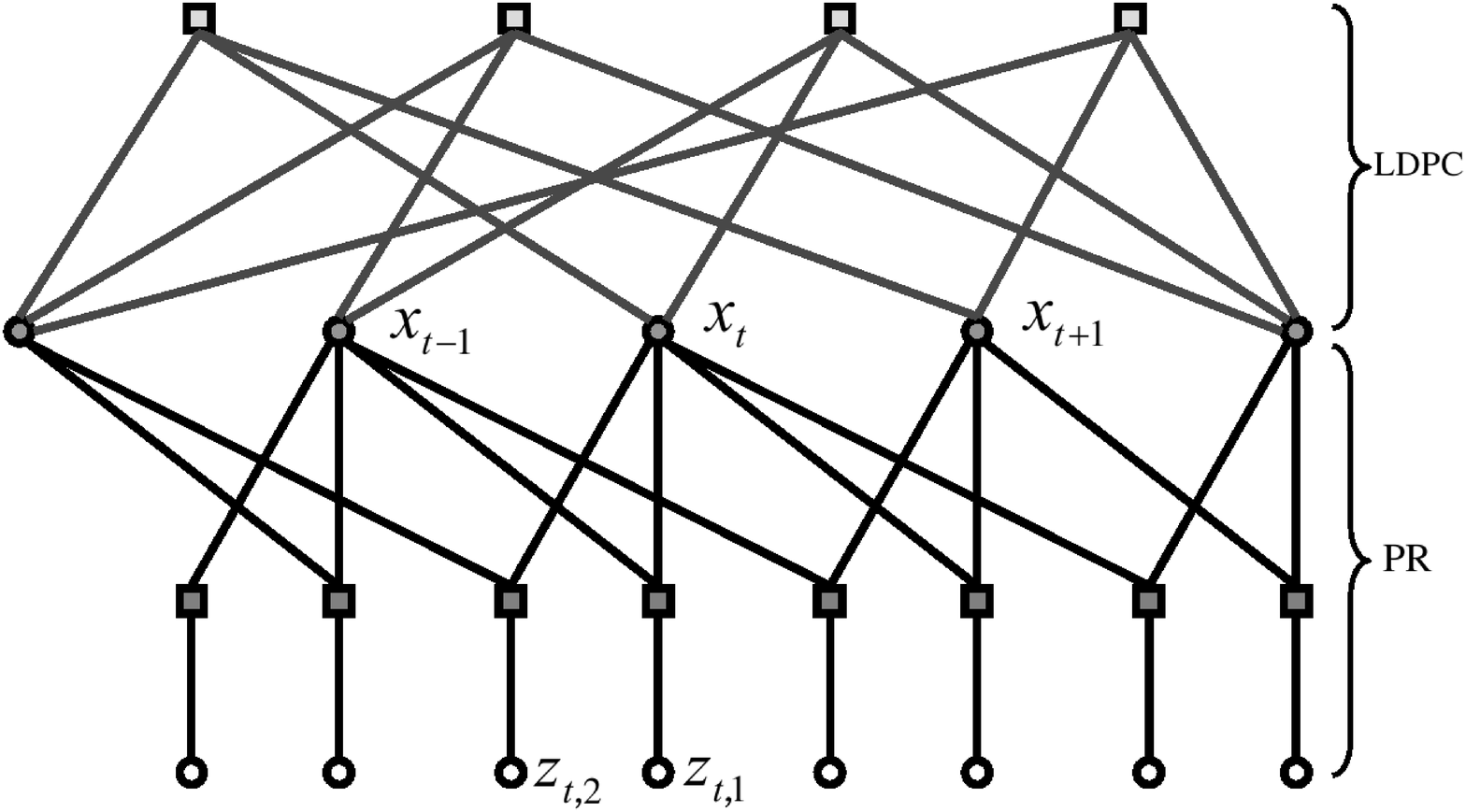}
\caption{PR channel and LDPC code represented by a Tanner graph.}\!\!\!\!\!\!\!\!\!\!\!
\label{PR-LDPC}
\end{figure}

The combined equalization and decoding problem (\ref{ILP}) has the form of a single decoding problem, which can be represented by a low-density Tanner graph. Fig. \ref{PR-LDPC} shows an example of the combination of a PR channel of memory size $2$ with an LDPC code. We call the upper and lower layers of this Tanner graph the code layer and the PR layer (or the PR graph), respectively. The PR layer of the graph consists of $\mu n$ check nodes $c_{t,j}$ of degree 3, each connected to two information bit nodes $x_t$, $x_{t-j}$, and one distinct state bit node, $z_{t,j}$. Also, the PR layer can contain cycles of length 6 and higher. If a coefficient, $\lambda_{t,j}$, is zero, its corresponding state bit node, $z_{t,j}$, and the check node it is connected to can be eliminated from the graph, as they have no effect on the decoding process. 

It follows from (\ref{lambda}) that the coefficients of the state bits in the objective function, $\{\lambda_{t,j}\}$, are only a function of the PR channel impulse response, while the coefficients of the information bits are the results of matched filtering the noisy received signal by the channel impulse response, and therefore dependent on the noise realization. Once the variable coefficients in the objective function are determined, LP decoding can be applied to solve a linear relaxation of decoding on this Tanner graph. We call this method \emph{LP detection}. In the relaxation of \cite{Feldman}, the binary parity-check constraint corresponding to each check node $c$ is relaxed as follows. Let $N_c$ be the index set of neighbors of check node $c$, i.e. the variable nodes it is directly connected to in the Tanner graph. Then, we include the following constraints
\begin{equation}
\label{constraints}
\sum_{i\in V} x_i -\sum_{i\in N_c\backslash V} x_i \leq |V|-1,\ \ \forall\ V\subset N_c\ \text{s.t.}\ |V|\ \text{is odd}.
\end{equation}
In addition, the integrality constraints $x_i\in \{0,1\}$ are relaxed to box constraints $0\leq x_i \leq 1$. This relaxation has the ``ML certificate property,'' i.e. if the solution of the relaxed LP is integral, it will also be the solution of (\ref{ILP}).

The coefficients in the linear objective function, after some normalization, can also be treated as log-likelihood ratios (LLR) of the corresponding bits, which can be used for iterative MP decoding. In this work, we have mostly used the Min-Sum Algorithm (MSA), since, similar to LP decoding, it is not affected by the uniform normalization of the variable coefficients in (\ref{ILP}). 
%Furthermore, one can argue that since each $q_t$ contains a Gaussian noise term with variance proportional to $\sigma^2$, a suitable normalization of the objective coefficients to estimate the ``equivalent LLRs'' is to divide them by the the variance of the noise, $\sigma^2$. In this work, we have also used the equivalent LLRs obtained by this normalization to perform decoding using the Sum-Product Algorithm (SPA).

\section{Performance Analysis of Uncoded Detection}
In this section, we study the performance of LP detection in the absence of coding, i.e. solving (\ref{MLD matrix form}) with $\mathscr{C}=\{0,1\}^n$. It is known that if the off-diagonal elements of $P$ are all nonpositive; i.e. $\lambda_{t,j}\geq 0,\ \forall j\neq0, t$, the 0-1 problem is solvable in polynomial time by reducing it to the MIN-CUT problem; e.g. see \cite{Rhys}. As an example, Sankaran and Ephremides \cite{MUD} argued using this fact that when the spreading sequences in a synchronous CDMA system have nonpositive cross correlations, optimal multiuser detection can be done in polynomial time. In this section, we derive a slightly weaker condition than the nonnegativity of $\lambda_{t,j}$, as the necessary and sufficient condition for the success of the LP relaxation to result in an integer solution for any value of $\underline{q}$ in (\ref{ILP}). This analysis also sheds some light on the question of how the algorithm behaves in the general case, where this condition is not satisfied.%, such as explaining why in some cases the algorithm fails to provide an integer solution even in the absence of noise.

%The Tanner graph of the PR channel has the property that all of its check nodes have degree 3. 
For a check node in the Tanner graph connecting information bit nodes $x_t$ and $x_{t-j}$ and state bit node $z_{t,j}$, the constraints (\ref{constraints}) can be summarized as
\begin{align}
\label{z-constraints}
z_{t,j} &\geq \text{max}[x_t-x_{t-j}, x_{t-j}-x_t] \nonumber\\
z_{t,j} &\leq \text{min} [x_t+x_{t-j}, 2-x_t-x_{t-j}],
\end{align}
which can be further simplified as
\begin{align}
\label{z-const}
|x_t-x_{t-j}| \leq z_{t,j} \leq 1-|x_t+x_{t-j}-1|.
\end{align}
Since there is exactly one such pair of upper and lower bounds for each state bit, in the solution vector, $z_{t,j}$ will be equal to either the lower or upper bound, depending on the sign of its coefficient in the linear objective function, $\lambda_{t,j}$. Hence, having the coefficients, the cost of $z_{t,j}$ in the objective function can be written as
\begin{equation}
\label{cost of z}
\lambda_{t,j} z_{t,j}= \begin{cases}
\lambda_{t,j} |x_t-x_{t-j}| & \text{if } \lambda_{t,j}\geq 0,\\
\lambda_{t,j} - \lambda_{t,j}|x_t+x_{t-j}-1| & \text{if } \lambda_{t,j} < 0,
\end{cases}
\end{equation}
where the first term in the second line is constant and does not affect the solution.
%Consequently, by projecting the LP into the original $n$-dimensional space, we will obtain the minimization problem
Consequently, by substituting (\ref{cost of z}) in the objective function, the LP problem will be projected into the original $n$-dimensional space, giving the equivalent minimization problem
\begin{align}
\label{convex problem}
\text{Minimize} \hspace{0.3 in} &f(\underline{x})=\sum_t q_t x_t + \mathop{\sum\sum}\limits_{t,j: \lambda_{t,j}>0} |\lambda_{t,j}| |x_t-x_{t-j}| \nonumber\\
& \hspace{0.4 in} + \mathop{\sum\sum}\limits_{t,j: \lambda_{t,j}<0} |\lambda_{t,j}||x_t+x_{t-j}-1|, \nonumber\\
% \text{Subject to} \hspace{0.3 in} &\underline{x} \in [0,1]^n,
\text{Subject to} \hspace{0.3 in} & 0 \leq x_t \leq 1,\ \forall t=1,\ldots, n,
\end{align}
which has a convex and piecewise-linear objective function. Each absolute value term in this expression corresponds to a check node in the PR layer of the Tanner graph representation of the channel. 
%Since (\ref{convex problem}) is equivalent to the LP relaxation of the ML detection problem, whenever it gives an integral solution, it will be the ML sequence. 

\subsection{LP-Proper Channels: Guaranteed ML Performance}
For a class of channels, which we call \emph{LP-proper channels}, the proposed LP relaxation of uncoded ML detection always gives the ML solution. The following theorem provides a criterion for recognizing LP-proper channels.
\begin{theorem}
\label{WNC}
The LP relaxation of the integer optimization problem (\ref{ILP}), in the absence of coding, is exact for every transmitted sequence and every noise configuration if and only if the following condition is satisfied for $\{\lambda_{t,j}\}$: 
\begin{quote}
\textbf{Weak Nonnegativity Condition (WNC)}: Every check node $c_{t,j}$, connected to variable nodes $x_t$ and $x_{t-j}$, which lies on a cycle in the PR Tanner graph corresponds to a nonnegative coefficient; i.e. $\lambda_{t,j} \geq 0$. 
\end{quote}
\end{theorem}

% \begin{remark}
% Satisfying WNC is a characteristic of the PR channel impulse response, and is independent of the transmitted data sequence and the noise configuration.
% \end{remark}

\begin{figure}
\centering
\includegraphics[width=2.5 in] {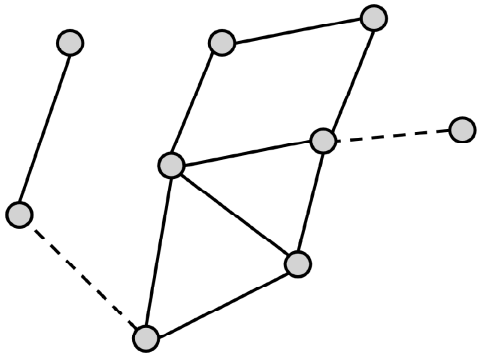}
\caption{The dependence graph of the system of linear equations with one cluster of cycles. Solid lines represent positive edges and dashed lines represent negative edges.}\!\!\!\!\!\!\!\!\!
\label{graph}
\end{figure}

\begin{proof}
We first prove that WNC is sufficient for guaranteed convergence of LP to the ML sequence, and then show that if this condition is not satisfied, there are cases where the LP algorithm fails. In the proof, we make use of the following definition.
\begin{definition} 
Consider a piecewise-linear function $f:\mathbb{R}^n \mapsto \mathbb{R}$.
We call $\underline{a}$ a \emph{breakpoint} of $f$ if the derivative of $f(\underline{a} + s \underline{v})$ with respect to $s$ changes at $s=0$, for any nonzero vector $\underline{v} \in \mathbb{R}^n$.
\end{definition}
\subsubsection{Sufficiency}
It is sufficient to show that under WNC, the solution of (\ref{convex problem}) is always at one of the vertices of the unit cube, $[0,1]^n$. First consider (\ref{convex problem}) without the box constraints. The optimum point, $\underline{x}^*$, of the objective function has to occur either at infinity or at a breakpoint of this piecewise-linear function. Since the nonlinearity in the function comes from the terms involving the absolute value function, each breakpoint, $\underline{a}$, is determined by making $n$ of these absolute value terms \emph{active}, i.e. setting their arguments equal to zero. These $n$ terms should have the property that the linear system of equations obtained by setting their arguments to zero has a unique solution.

When the feasible region is restricted to the unit cube $[0,1]^n$, the optimum can also occur at the boundaries of this region.
% If this optimum point in one of the vertices of the cube, then an integral solution is found. 
Without loss of generality, we can assume that the optimum point lies on a number, $k$, of hyperplanes corresponding to the box constraints in (\ref{convex problem}), where $k=0,\ldots,n$. This will make exactly $k$ variables, $x_i, \ i\in I$, equal to either 0 or 1, where $|I|=k$. In addition, at least $n-k$ other equations are needed to determine the remaining $n-k$ fractional variables. These equations will be the results of making a number of absolute value terms active, each of them having one of the two forms $x_t=x_{t-j}$ or $x_t+x_{t+j}=1$, depending on whether $\lambda_{t,j}>0$ or $\lambda_{t,j}<0$, respectively. When an absolute value term in (\ref{convex problem}) is active, either both, or none of its variables can be integer. Since the former case does not provide an equation in terms of the fractional variables, we can assume that all these active absolute value terms only involve fractional variables. 

Now the question becomes under what condition such equations can have a unique and nonintegral solution. We can illustrate this system of equations by a \emph{dependence graph}, where the vertices correspond to the unknowns, i.e., the $n-k$ fractional variable nodes, and between vertices $x_s$ and $x_{s-i}$ there is a \emph{positive} edge if $\lambda_{s,i}>0$ and a \emph{negative} edge if $\lambda_{s,i}<0$. An example of a dependence graph satisfying WNC is shown in Fig. \ref{graph}. 
% In general, this graph contains $L$ clusters, $C_1, C_2, \ldots, C_L$, of cycles, and also $J$ additional edges which do not lie on any cycle. If WNC is satisfied, each cluster will contain only positive edges. This means that if we separate the equations corresponding only to vertices within a cluster $C_i$ of cycles, these equations will all have the form $x_t=x_{t-j}$, where vertices $t$ and $t-j$ belong to $C_i$. This means that all vertices in $C_i$ have the same value, $\beta_i$, but $\beta_i$ cannot be determined by these equations. The values of $\{\beta_i\}$ should be determined by the edges (equations) between the clusters. If we modify the dependence graph by concentrating each cluster to a vertex of value $\beta_i$, the graph will not contain any cycles. Therefore, the system of equations for determining the variables is under-determined, and none of the nodes in this graph will have a unique solution. Hence, the only case that will have a unique solution for all the variables will be $k=n$, which means that all of the variables $\{x_i\}$ are integral. This proves the sufficiency of WNC.
In the solution of the system of equations, if two vertices are connected in the dependence graph by a positive edge, they will have the same value. Hence, we can merge these two vertices into a single vertex, and the value that this vertex takes will be shared by the two original vertices. If we do this for every positive edge, we will be left with a reduced dependence graph that has only negative edges. We claim that, if WNC is satisfied, the reduced graph will be tree. To see this, consider a negative edge $e_{t,j}$ connecting vertex $x_t$ on its ``left side'' to vertex $x_{t-j}$ on its ``right side'' in the original dependence graph. By assumption, $e_{t,j}$ is not on a cycle, which means if we remove it, its left side and right side will become disconnected. Clearly, during the above-mentioned merging of vertices, no new connection will be created between the left and right sides of $e_{t,j}$. Hence, every negative edge will still not be on any cycle at the end of the merging procedure, and, in other words, the reduced dependence graph will be a tree. Since trees have fewer edges than vertices, the system of equations for determining the unknown variables will be under-determined, and none of the nodes in the dependence graph will have a unique solution. Consequently, the only case where we have a unique solution for all the variables will be $k=n$, which means that all of the variables $\{x_i\}$ are integral. This proves the sufficiency of WNC.

\subsubsection{Necessity}
We prove the necessity of the condition by a counter example. Consider a case where the realization of the noise sequence is such that the received sequence is zero. This will make $\{q_t\}$, the coefficients of the linear term in (\ref{convex problem}), equal to zero. Hence we are left with the positive-weighted sum of a number of absolute value terms, each of them being greater than or equal to zero. The objective function will become zero if and only if all these terms are zero, which is satisfied if $\underline{x} = [\frac{1}{2}, \ldots, \frac{1}{2}]^T$. We need to show that if WNC is not satisfied, equating all the absolute value terms to zero will determine a unique and fractional value for at least one of the elements of $\underline{x}$. Consider the dependence graph of this system of equations. We know that there is at least one cycle containing a negative edge. Consider the equations corresponding to one such cycle. Without loss of generality, we can assume that all these equations have the form $x_t+x_{t+j}=1$, since if any of them has the form $x_t=x_{t+j}$, we can combine these two variables into one. Consequently, the system of equations corresponding to this cycle, after some column permutations will have the matrix form
\begin{equation}
\label{linear equations}
\begin{bmatrix}
1 & 1 & 0 &\cdots\\
0 & 1 & 1 & 0\\
\vdots & &  \ddots\\
1 & \cdots & 0 & 1 
\end{bmatrix}
\begin{bmatrix}
x_{i_1}\\
x_{i_2}\\
\vdots\\
x_{i_l}
\end{bmatrix}
= 
\begin{bmatrix}
1\\
1\\
\vdots\\
1
\end{bmatrix}
.
\end{equation}
Since the coefficient matrix is full rank, the unique solution of this system is $x_{i_k}= \frac{1}{2}, k=1,\ldots, l$. This means that no integral vector $\underline{x}$ can make the objective function in (\ref{convex problem}) zero, and therefore the algorithm fails to find an integral solution. This proves the necessity of WNC for guaranteed success of the LP relaxation.
\end{proof}

\begin{corollary}
\label{fractional}
The solutions of the LP relaxation of uncoded ML equalization are in the space $\{0, \frac{1}{2}, 1\}^n$.
\end{corollary}
\begin{proof}
The values of fractional elements of $\underline{x}$ are the unique solutions of a system of linear equations of the forms $x_t=x_{t+i}$ and $x_t+x_{t+i}=1$. The vector $[\frac{1}{2}, \ldots, \frac{1}{2}]$ satisfies all these equations, and, hence, has to be the unique solution.
\end{proof}

\subsection{Implications of the WNC}
In the PR channel equalization problem, due to the periodic structure of the Tanner graph and the coefficients of the state variables, the WNC implies that at least one of the following statements should be valid:
\begin{enumerate} 
\item The PR Tanner graph is acyclic.
%, or equivalently, there is only one nonzero coefficient $\lambda_{t,j}$ for each $t$. 
Examples include any PR channel with memory size $\mu=1$, and the memory-2 channel $h(D)=1+D-D^2$.
\item \textbf{Nonnegativity Condition (NC):} All state variables have nonnegative coefficients; i.e. $\lambda_{t,j}\geq 0\ \forall t,j$.
\end{enumerate}

\begin{lemma}
Condition 1 implies that, among $\{\lambda_{t,j}:\ j=1,2,\ldots$\}, on average at most one can be nonzero for each $t$. 
\end{lemma}
\begin{proof}
Let $\kappa$ be the average number of nonzero elements of $\{\lambda_{t,j}:\ j=1,2,\ldots$\} for each $t$. Then, it is easy to see that the PR Tanner graph will have $3\kappa n$ edges and $n(1+2\kappa)$ vertices (i.e. variable or check nodes). But the graph can be acyclic only if $3\kappa n \leq n(1+2\kappa)-1$, which means that $\kappa < 1$.
\end{proof}

In a one-dimensional ISI channel, where the state coefficients are given by (\ref{lambda}), NC implies that the autocorrelation function of the discrete-time impulse response of the channel should be nonpositive at any point other that the zero time shift. As the memory size of the channel increases, this condition becomes more restrictive, so that a long and randomly-chosen impulse response is not very likely to satisfy NC. However, it is very common, particularly in magnetic recording applications, to use the PRML technique, where the overall impulse response of the channel is first equalized to a target impulse response, in order to simplify the subsequent detection stages. The target channel response is usually optimized to provide the best detection performance. A possible approach for employing the linear relaxation of ML detection in this application can be to optimize the target channel subject to NC, which enables us to achieve the performance of Viterbi detection, although without an exponential complexity in the memory size.

An interesting application of this result is in 2-D channels, for which there is no feasible extension of the Viterbi algorithm. In a 2-D channel, the received signal $r_{t,s}$ at coordinate $(t,s)$ in terms of the transmitted symbol sequence $\{\tilde{x}_{t,s}\}$ has the form
\begin{equation}
\label{2D}
r_{t,s}=\sum_{i=0}^{\mu}\sum_{j=0}^{\nu} h_{i,j} \tilde{x}_{t-i,s-j} +n_{t,s}.
\end{equation}
Hence, following the same procedure that results in (\ref{lambda}), the coefficients of the state variables in the 2-D channel can also be obtained. In particular, the state variable defined as $z_{(t,s),(k,l)}= x_{t,s} \oplus x_{t-k,s-l}$ will have the coefficient
\begin{equation}
\label{2D lambda}
\gamma_{k,l}=-\sum_{i=0}^{\mu}\sum_{j=0}^{\nu} h_{i,j} h_{i+l,j+l}.
\end{equation}
In this expression, for simplicity, we have dropped the $(t,s)$ index due to the independence of the $\gamma$ from $t$ and $s$, except near the boundaries, which, in turn, can be resolved by proper truncation of the transmitted array. Theorem \ref{WNC} guarantees that ML detection can be achieved by linear relaxation if NC is satisfied, i.e. $\gamma_{k,l}\geq 0$ for any $k,l>0$. An example of a 2-D channel satisfying NC is given by the matrix
\begin{equation}
\label{2D example}
[h_{i,j}]=
\begin{bmatrix}
1& \ 1\\
1& -1
\end{bmatrix}
.
\end{equation}

\subsection{Error Event Characterization: asymptotically LP-Proper and LP-Improper Channels}
We showed that if WNC is not satisfied, there are noise configurations that results in the failure of LP detection to find the ML sequence. However, for some channels, such noise configurations might be highly unlikely at high SNR values. We have observed that for some ISI channels violating WNC, the probability of obtaining a fractional solution becomes dominated by the probability that the ML sequence is different from the transmitted word. We call these channels \emph{asymptotically LP-proper}, since for these channels the WER of LP detection is asymptotically equal to that of ML detection as the SNR increases. On the other hand, for some other channels, which we call \emph{LP-improper} channels, there is a non-diminishing probability that the solution of LP detection is nonintegral, as the SNR increases. In this subsection, we study the performance of the uncoded LP detection for general ISI channels and derive conditions for failure of the detector to find the ML solution. These conditions will help us to classify different channels based on their performance. With some modifications, these conditions can also be applied to 2-D ISI channels. Since here we focus on stationary 1-D PR channels, as explained in Section II, we can assume that $\lambda_{t,j} = \lambda_j$ is independent of $t$.

\begin{definition}
\label{graph def}
Given the solution of LP detection, the \emph{fractional set}, $F=\{i_1,\ldots,i_{n-k}\}\subset \{1,\ldots,n\}$, is the set of indices of information bit nodes in the Tanner graph of the PR channel that have fractional values in the solution, $\underline{\hat{x}}$. 
% The \emph{fractional subgraph} $\Phi$ is the subgraph of the Tanner graph containing these bit nodes, all their adjacent check nodes, and all the edges connecting these bit and check nodes. $\Phi$ can contain a number of disjoint, connected subgraphs, $\Phi_1, \ldots, \Phi_K$. We denote by $F_i$ the index set of information bit nodes in $\Phi_i$.
\end{definition}

We know from Corollary \ref{fractional} that the fractional elements in the solution, $\underline{\hat{x}}$, are all equal to $\frac{1}{2}$. A reasonable assumption supported by our simulations at high SNR is that if the ML solution, $\underline{x}$, is correct, the integer elements of $\underline{\hat{x}}$ are correct, as well. In other words, we have
\begin{equation}
\label{x and xhat}
\hat{x}_i= \begin{cases}
\frac{1}{2} & \text{if } i\in F,\\
x_i & \text{if } i\notin F.
\end{cases}
\end{equation}

For the objective function $f(\cdot)$ of (\ref{convex problem}) we can write 
\begin{equation}
\label{g(x,x_hat)}
g(\underline{x} , \underline{\hat{x}}) \triangleq f(\underline{x}) - f(\underline{\hat{x}}) \geq 0.
\end{equation}
By expanding $f$ using (\ref{convex problem}), this inequality can be written in terms of $\{\lambda_{t,j}\}$, $\underline{x}$, and $\underline{q}$. Before doing so, we present the following lemma to simplify the absolute value terms in (\ref{convex problem}).

\begin{lemma}
\label{simplify}
Let $x_t$ and $x_s$ be binary variables and $\tilde{x}_t$ and $\tilde{x}_s$ be their bipolar versions, respectively. In addition, let's define
\begin{equation}
%h(x_t,x_s,\lambda) \triangleq 
%\begin{cases}
%|\lambda_{|t-s|}| |x_t-x_{t-j}| & \text{if } \lambda_{|t-s|}>0,\\
%|\lambda_{|t-s|}| |x_t+x_{t-j}-1| & \text{if } \lambda_{|t-s|}<0.
%\end{cases}
%\end{equation}
h(x,y,\lambda) \triangleq 
\begin{cases}
|\lambda| |x-y| & \text{if } \lambda \geq 0,\\
|\lambda| |x+y-1| & \text{if } \lambda<0.
\end{cases}
\end{equation}
Then, the following equations hold:
\begin{enumerate}
\item $h(x_t, x_s, \lambda_{|t-s|}) = \frac{1}{2} |\lambda_{|t-s|}| - \frac{1}{2} \lambda_{|t-s|} \tilde{x}_t \tilde{x}_s,$
\item $h(x_t, \frac{1}{2}, \lambda_{|t-s|}) = \frac{1}{2} |\lambda_{|t-s|}|,$
\item $h(\frac{1}{2}, \frac{1}{2}, \lambda_{|t-s|})=0.$
\end{enumerate}
\end{lemma}

\begin{proof}
The equations can be verified by using (\ref{bipolar}), and checking the possible values for $x_t$ and $x_s$.
\end{proof}

By using Lemma \ref{simplify}, and cancelling the terms that are common between $f(\underline{x})$ and $f(\underline{\hat{x}})$, $g(\underline{x} , \underline{\hat{x}})$ can be written as
\begin{align}
\label{expand g}
g(\underline{x} , \underline{\hat{x}})= \sum \limits_{t\in F} \bigg[ q_t \big(x_t-\frac{1}{2} \big) 
+ \sum \limits_{s\in F,\ s<t} \Big( \frac{1}{2} |\lambda_{|t-s|}| - \frac{1}{2} \lambda_{|t-s|} \tilde{x}_t \tilde{x}_s \Big)
- \sum \limits_{s\notin F} \frac{1}{2} \lambda_{|t-s|} \tilde{x}_t \tilde{x}_s \bigg],
\end{align}
where $\lambda_{d} = -\sum_i h_i h_{i+d}$ is defined to be zero if $d>\mu$. Since we assumed that $\underline{x}$ is equal to the transmitted sequence, we can expand $q_t$ as
\begin{align}
\label{expand q_t}
q_t &= \sum\limits_{i=0}^{\mu} h_i r_{t+i} \nonumber \\
&= \sum\limits_{i=0}^{\mu} \sum\limits_{i=0}^{\mu} h_i h_j \tilde{x}_{t+i-j} + \sum\limits_{i=0}^{\mu} h_i n_{t+i} \nonumber \\
&= -\sum\limits_{s= t-\mu}^{t+\mu} \lambda_{|t-s|} \tilde{x}_s + \eta_t,
\end{align}
where $\eta_t \triangleq \sum_i h_i n_{t+i}$. By substituting (\ref{expand q_t}) into (\ref{expand g}), and using the fact that $x_t - \frac{1}{2} = -\frac{1}{2} \tilde{x}_t$, we obtain

\begin{align}
\label{simplify g}
g(\underline{x} , \underline{\hat{x}}) & =\sum \limits_{t\in F} \bigg[ \frac{1}{2}\lambda_0 + \sum \limits_{s\in F,\ s<t} \Big( \frac{1}{2} |\lambda_{|t-s|}| + \frac{1}{2} \lambda_{|t-s|} \tilde{x}_t \tilde{x}_s \Big) - \frac{1}{2} \eta_t \tilde{x}_t \bigg] \nonumber\\
& = \frac{1}{2} \bigg[ c_F + \frac{1}{2} \underline{\tilde{x}}^T_F \bar{P}_F \underline{\tilde{x}}_F + \underline{\eta}^T_F \underline{\tilde{x}}_F \bigg].
\end{align}
In this equation, $\underline{\tilde{x}}_F$ and $\underline{\eta}_F$ are obtained respectively from $\underline{\tilde{x}}$ and $\underline{\eta}$ by keeping only the elements with indices in $F$, $\bar{P}_F$ is a submatrix of $P$ (defined in II-B) consisting of the elements of $P$ with column and row indices in $F$ and its diagonal elements made equal to zero, and
\begin{equation}
\label{c_F}
c_F \triangleq \sum \limits_{t\in F} \lambda_0 + \mathop{\sum \sum}\limits_{t,\,s\in F,\ s<t} |\lambda_{|t-s|}|.
\end{equation}
Equations (\ref{g(x,x_hat)}) and (\ref{simplify g}) lead us to the following Theorem.
\begin{theorem}
\label{failure condition}
Uncoded LP detection fails to find the transmitted sequence if there is an index set $F\subset \{1,\ldots,n\}$ for which
\begin{equation}
\label{suff failure}
c_F + \frac{1}{2} \underline{\tilde{x}}^T_F \bar{P}_F \underline{\tilde{x}}_F + \underline{\eta}^T_F \underline{\tilde{x}}_F > 0. 
\end{equation}$\hfill\QED$
\end{theorem}

If the transmitted sequence, $\underline{x}$, is given, we can estimate the probability that the sufficient failure condition given by Theorem \ref{failure condition} is satisfied, and determine the dominant error event causing this failure. In order to do that, for any given $F$, we can calculate a ``distance'' for the error event corresponding to $F$ defined as
\begin{equation}
\label{d_F}
d_F = - c_F - \frac{1}{2} \underline{\tilde{x}}^T_F \bar{P}_F \underline{\tilde{x}}_F,
\end{equation}
and the variance of the noise corresponding to this error event, given by
\begin{align}
\label{sigma2_F}
\sigma^2_F &\triangleq \text{var} \Big[ \underline{\eta}^T_F \underline{\tilde{x}}_F\Big] \nonumber\\
& = \text{var} \bigg[ \sum\limits_s n_s \sum\limits_{t=s-\mu,\ t\in F}^s \tilde{x}_t h_{s-t} \bigg] \nonumber\\
& = \sigma^2 \sum\limits_s \Big[ \sum\limits_{t=s-\mu,\ t\in F}^s \tilde{x}_t h_{s-t} \Big]^2,
\end{align}
where $\sigma^2$ is the variance of each noise sample, $n_t$. Hence, the probability that the error event corresponding to $F$ occurs will be equal to 
\begin{equation}
\label{prob(F)}
p(F,\underline{\tilde{x}}_F) = Q(\frac{d_F}{\sigma_F}),
\end{equation}
where $Q(x)$ is the Gaussian Q function. 

In order to find the dominant error event over all transmitted sequences, for every choice of the index set, $F$, we should find the vector $\underline{\tilde{x}}_F \in \{-1,\, 1\}^{|F|}$ that maximizes the probability in (\ref{prob(F)}). However, this will require an exhaustive search over all $\underline{\tilde{x}}_F$. As an alternative, we can upper bound this probability by finding the smallest distance $d_F^{\min} \triangleq \min_{\underline{\tilde{x}}_F} d_F$ and the largest variance ${\sigma^2_F}^{\max} \triangleq \max_{\underline{\tilde{x}}_F} \sigma^2_F$, and computing $Q(d_F^{\min} / \sigma_F^{\max})$. Fortunately, each of these two optimization problems can be solved by dynamic programming (Viterbi algorithm) over a trellis of at most $2^\mu$ states.

Specifically, if the minimum distance is negative, there is a probability independent of the SNR that the sequence $\underline{\tilde{x}}_F$ corresponding to that distance exists in the transmitted block, and given that event, the probability of failure, $(\ref{prob(F)})$, will be greater than $\frac{1}{2}$ for any SNR value. Therefore, there will be a non-diminishing probability of failure as SNR goes to infinity. This motivates the following results:
\begin{corollary}[LP-Improper Channels]
If for an ISI channel, there is a index set $F \subset \{1,\ldots,n\}$ and a vector $\underline{\tilde{x}}_F \in \{-1,\, 1\}^{|F|}$ for which $d_F$ defined in (\ref{d_F}) is negative, LP detection on this channel will have a non-diminishing WER as SNR grows; i.e., the channel is \emph{improper}. $\hfill\QED$
\end{corollary}

On the other hand, if for an ISI channel the probability of failure computed by the proposed technique decreases more steeply than the WER of ML detection as SNR increases, the WER of LP detection will be asymptotically equal to that of ML detection. In this case, the channel will be \emph{asymptotically LP-proper}.

\begin{remark}
The error events considered in this analysis are not the only possible types of detector failure. This analysis is intended to approximate the gap between the performance of LP and ML detection methods by estimating the probability that LP detection fails to find the integer optimum (i.e., ML) solution, given the ML detector is successful. As mentioned before, we only studied the events where a vector having the form of (\ref{x and xhat}) has a lower cost than the transmitted word. Therefore, even if (\ref{suff failure}) does not hold for any $F$, it is theoretically possible, although not very likely, as we observed in practice, that LP detection has a fractional solution.
\end{remark}

Of particular interest among the possible error events is the one where the all-$\frac{1}{2}$ vector has a lower cost than the correct solution; i.e., $F=\{1,\ldots,n\}$ in (\ref{suff failure}). For a given transmitted sequence $\underline{x}$, this event is not necessarily the most likely one. However, studying this event provides us with a simplified sufficient condition for the failure of LP detection, which further clarifies the distinction between the different classes of ISI channels.

The distance, $\delta$, corresponding to this event is obtained by putting $F=\{1,\ldots,n\}$ in (\ref{d_F}). If the block length, $n$ is much larger than the channel memory length, $\mu$, we can neglect the ``edge effects'' caused by the indices that are within a distance $\mu$ of one of the two ends of the sequence; thus, we will have
\begin{align}
\label{all-1/2 distance}
\delta &= -n\lambda_0 - n\sum\limits_{j=1}^{\mu} |\lambda_j| - \sum\limits_{j=1}^{\mu} \lambda_j \sum\limits_t \tilde{x}_t \tilde{x}_{t+j} \nonumber\\
&= -n\lambda_0 - n\sum\limits_{j=1}^{\mu} |\lambda_j| - \sum\limits_{j=1}^{\mu} \lambda_j \rho_{j},
\end{align}
where $\rho_j$ is the autocorrelation of $\underline{\tilde{x}}$ with a shift equal to $j$. On the other hand, for the noise variance, $\varsigma^2$, corresponding to this event we have from (\ref{sigma2_F})
\begin{align}
\label{all-1/2 variance}
\varsigma^2 &= \sigma^2 \sum\limits_t \bigg[ \sum\limits_{j=0}^\mu h_j \tilde{x}_{t-j} \bigg]^2 \nonumber\\
&= \sigma^2 \underline{\tilde{x}}^T P \underline{\tilde{x}} \nonumber\\
&= -\sigma^2 n \lambda_0 - 2\sigma^2 \sum\limits_{j=1}^{\mu} \lambda_j \rho_{j}.
\end{align}
Note that $\delta$ and $\varsigma^2$ have a similar dependence on the transmitted sequence. 

A possible approach for finding the likelihood of occurrence of an all-$\frac{1}{2}$ error event is to maximize $\frac{\delta}{\varsigma}$ over all possible transmitted sequences. However, with a random transmitted sequence, the probability that this quantity becomes close to its worst case may become very low for long block lengths. As an alternative, here we show that as $n$ grows while $\mu$ remains fixed, the dependence of both $\delta$ and $\varsigma^2$ on the transmitted sequence become negligible compared to the constant term. 

\begin{lemma}
\label{bound rho}
Let $\tilde{x}_1, \ldots , \tilde{x}_n $ be a sequence of i.i.d. $\pm 1$ random variables, each equally likely to be $+1$ or $-1$, and let $\rho_j=\sum_t^{n-j} \tilde{x}_t \tilde{x}_{t+j}$. Then, for fixed $\mu$, as $n \to \infty$
\begin{equation}
%\Pr \bigg[ \Big|\sum\limits_{j=1}^{\mu} \lambda_j \rho_{j} \Big| > \mu |\lambda_0| n^{\frac{1}{2}+\epsilon} \bigg] \to 0\ \ \text{as } n \to \infty.
\frac{1}{n} \sum\limits_{j=1}^{\mu} \lambda_j \rho_{j} \to 0\ \ \text{almost surely}.
\end{equation}
\end{lemma}
\begin{proof}
For each $j=1,\ldots,\mu$, $\rho_j$ is the sum of $n-j$ terms of the form $\tilde{x}_t \tilde{x}_{t+j}$. Clearly, each of these terms is equally likely to be equal to $+1$ or $-1$. Furthermore, it can be shown that these terms are mutually independent\footnote{For a proof of this statement refer to Proposition 1.1 of \cite{Mercer}.}. Hence, using the strong law of large numbers, we have
\begin{align}
\frac{\rho_j}{n} &= \frac{n-j}{n}\Big( \frac{1}{n-j}\sum\limits_{t=1}^{n-j} \tilde{x}_t \tilde{x}_{t+j} \Big) \to 0\ \ \text{almost surely},
\end{align}
where we used the fact that ${n-j}/{n} \to 1$, since $1\leq j \leq \mu$. Consequently, 
\begin{align}
\frac{1}{n} \sum\limits_{j=1}^{\mu} \lambda_j \rho_{j} = \sum\limits_{j=1}^{\mu} \lambda_j \frac{\rho_{j}}{n} \to 0\ \ \text{almost surely},
\end{align}
since it is a linear combination of a finite number of variables, each going to zero almost surely.
\end{proof}

Using Lemma \ref{bound rho} in (\ref{all-1/2 distance}) and (\ref{all-1/2 variance}) for large $n$, we can write
\begin{equation}
\label{d_1/2}
\delta = n \Big[ |\lambda_0| - \sum\limits_{j=1}^{\mu} |\lambda_j| + o(1) \Big],
\end{equation}
and
\begin{equation}
\label{sigma^2_1/2}
\varsigma^2 = \sigma^2 n \big[|\lambda_0| + o(1) \big],
\end{equation}
where we used the fact that $\lambda_0 = -\sum h_i^2 \leq 0$. Since the probability of the all-$\frac{1}{2}$ error event is equal to $Q(\delta/\varsigma)$, the above results motivate us to define the following parameter to characterize this probability in the limit of $n\to \infty$.

\begin{definition}
The \emph{LP distance}, $\delta_\infty$, of a partial-response channel is given by
\begin{equation}
\label{define d_inf}
\delta_\infty = \frac{1}{|\lambda_0|} \Big(|\lambda_0| - \sum\limits_{j=1}^{\mu} |\lambda_j|\Big).
\end{equation}

The LP distance is dimention-less parameter that can take values between $-\infty$ and $1$. The following theorem gives a new sufficient condition in terms of the LP-distance for a channel to be LP-improper.
\end{definition}
\begin{theorem}
\label{infinite improper}
The WER of uncoded LP detection over an ISI channel with the transmitted sequence generated as a random sequence of i.i.d. Bernouli$(1/2)$ binary symbols goes to 1 as the block length $n$ goes to infinity for any SNR, i.e., the channel is LP-improper, if the LP distance, $\delta_\infty$, of the channel is nonpositive
% ; i.e.
% \begin{equation}
% \label{def d_inf}
% \delta_\infty \triangleq \frac{1}{|\lambda_0|} \Big(|\lambda_0| - \sum\limits_{j=1}^{\mu} |\lambda_j|\Big) < 0.
% \end{equation}
\end{theorem}

\begin{proof}
As mentioned earlier, the probability of the all-$\frac{1}{2}$ event is equal to $Q(\delta/\varsigma)$. From (\ref{d_1/2}) and (\ref{sigma^2_1/2}), for large $n$ we have
\begin{align}
\label{limit of d/sigma}
% \lim\limits_{n \to \infty} \frac{\delta}{\varsigma} = \lim\limits_{n \to \infty} \delta_{\infty}\frac{\sqrt{|\lambda_0|} \sqrt{n}}{\sigma}.
\frac{\delta}{\varsigma} = \sqrt{n} \Big( \delta_{\infty}\frac{\sqrt{|\lambda_0|} }{\sigma} + o(1)\Big).
\end{align}
If $\delta_\infty <0$, the right-hand side will approach $-\infty$ as $n$ increases, hence, $Q(\delta/\varsigma)$ will go to 1.
\end{proof}

Having $\delta_\infty$ as a measure of LP-properness for LP detection, it is interesting to study how it behaves for LP-proper channels; i.e., those satisfying WNC in Theorem \ref{WNC}. The following lemma provides an answer to this question.

\begin{lemma}
\label{d_inf for proper}
For LP-proper channels that satisfy NC (defined in in III-B) $\delta_\infty > \frac{1}{2} |\lambda_0|$.
\end{lemma}

\begin{proof}
For any ISI channel, we can write
\begin{align}
\label{expand sum h_i}
\Big[\sum\limits_{i=0}^\mu h_i \Big]^2 &= \sum\limits_i {h_i^2} + \mathop{\sum\sum}\limits_{i,j;\ i\neq j} h_i h_j \nonumber\\
&= |\lambda_0| - 2 \sum\limits_{j=1}^\mu \lambda_j \geq 0.
\end{align}
%If WNC is satisfied, as discussed in III-B, one of the following should hold:
%\begin{enumerate}
%\item Among all $j\geq 1$, there is only on $j^*$ such that $\lambda_{j^*} \neq 0$. 
%\item $\lambda_j \geq 0,\ \forall j\geq 1$.
%\end{enumerate}
%In the first case, from (\ref{expand sum h_i}) we have
%\begin{align}
%\lambda_k \leq \frac{|\lambda_0|}{2},
%\end{align}
%which yields
%\begin{align}
%\delta_\infty &= |lambda_0|-|lambda_k|
%&\geq 
%\end{align}
%In the second case, 
Hence, 
\begin{equation}
\sum \limits_{j=1}^\mu \lambda_j \leq \frac{1}{2}|\lambda_0|.
\end{equation}
Since NC is satisfied, $\lambda_j \geq 0,\ \forall j\geq 1$. Therefore, we have
\begin{align}
\delta_\infty &= |\lambda_0| - \sum\limits_{j=1}^{\mu} \lambda_j \nonumber\\
& \geq \frac{1}{2} |\lambda_0|.
\end{align}

\end{proof}

\subsection{Simulation Results}
\begin{figure}
\centering
\includegraphics[width=4.0 in] {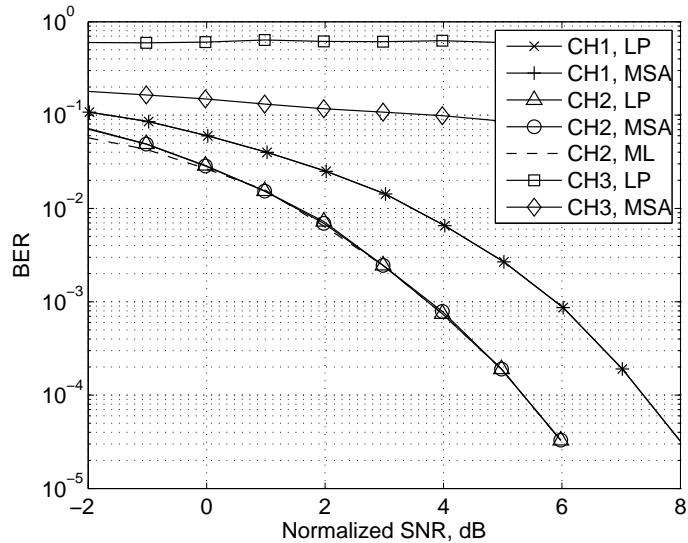}
\caption{BER for CH1-CH3. SNR is defined as the transmitted signal power to the received noise variance.}\!\!\!\!\!\!\!\!\!
\label{uncodedBER}
\end{figure}

We have simulated channel detection on the PR Tanner graph using LP decoding and MSA for three PR channels of memory size 3: 
\begin{enumerate}
\item \textbf{CH1:} $h(D)=1-D-0.5D^2-0.5D^3$ (with $\delta_\infty=\frac{1}{2}$, satisfies WNC; LP-proper),
\item \textbf{CH2:} $h(D)=1+D-D^2+D^3$ (with $\delta_\infty=\frac{1}{2}$),
\item \textbf{CH3:} $h(D)=1+D-D^2-D^3$ (with $\delta_\infty=0$; LP-improper). 
\end{enumerate}
Uncoded bit error rates (BER) of detection on these channels using LP and MSA are shown Fig. \ref{uncodedBER}. Since CH1 satisfies WNC, LP will be equivalent to ML on this channel. For CH2, we have also provided the BER of ML. Except at very low SNR where we see a small difference, the performance of LP and ML are nearly equal, which means that CH2 is an asymptotically LP-proper channel.
For both CH1 and CH2, MSA, converges in at most 3 iterations and has a BER very close to that of LP. On the other hand, for CH3, which is an LP-improper channel, we observe that the BERs of LP and MSA are almost constant. 
%The results for detection with LDPC coding are omitted due to space limitations. However, for the cases that we have simulated, all the LP-proper and asymptotically LP-proper channels also showed a good performance in the presence of coding, while LP-improper channels had a non-diminishing error rate at all SNR values. 

\begin{figure}
\centering
\includegraphics[width=4.0 in] {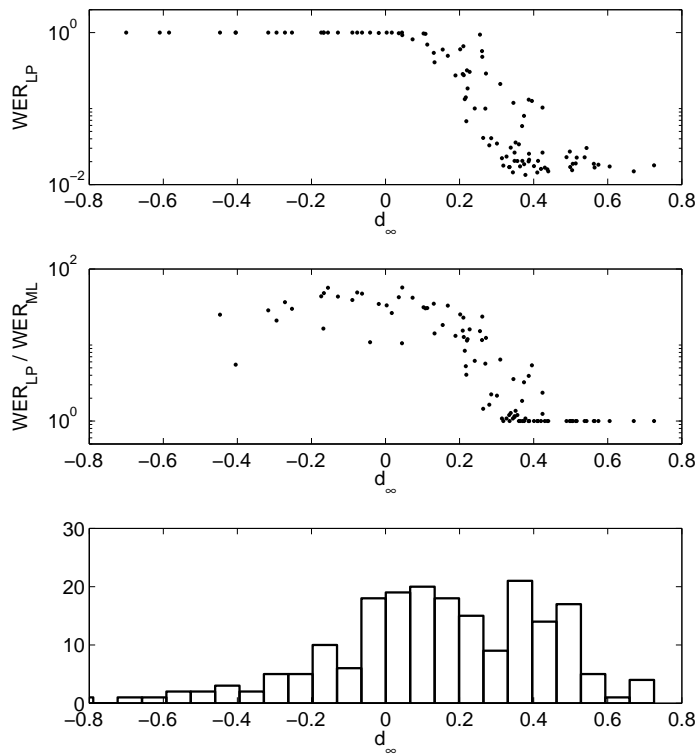}
\caption{Upper and middle plots: Performance of LP detection versus $\delta_\infty$ for random ISI channels of memory 4 at the SNR of 11 dB. Lower plot: The histogram of $\delta_\infty$.}\!\!\!\!\!\!\!\!\!
\label{d_inf}
\end{figure}

In Fig. \ref{d_inf}, we studied the effect of $\delta_\infty$ of ISI channels on the performance of LP detection. In this scenario, we randomly generate 200 ISI channels with memory 4, such that the taps of the impulse response are i.i.d. with a zero-mean Gaussian distribution. In addition, we normalize each channel so that the total energy of the impulse response is one; i.e. $|\lambda_0|=1$. We simulated the uncoded LP detection with random transmitted sequences of length 100 at the SNR of 11 dB. In the upper and middle plots of Fig. \ref{d_inf}, respectively, the WER of LP detection and the ratio of the WER of LP detection to that of ML detection are shown versus $\delta_\infty$ for these channels. In this work, ML detection was performed by using the cutting-plane method proposed in \cite{adaptive LP}. This method is based on using redundant parity checks (RPC) generated by modulo-2 combination of a subset of parity-check constraints. Once the LP decoding results in a nonintegral solution, we look for RPCs that, after linear relaxation, introduce a cut, i.e., make the current solution infeasible, and re-solve the LP after adding these constraints. This is continued until we obtain an integral solution, which is the ML sequence. For 192 of the 200 channels that we studied, this algorithm always successfully provided the ML solution with a few iterations.

The results in Fig. \ref{d_inf} demonstrate a strong correlation between the performance of the algorithm and the value of $\delta_\infty$. In particular, almost every channel with $\delta_\infty < 0.1$ has a WER close to 1, while for almost every channel with $\delta_\infty > 0.4$, the WER of LP is very close to that of ML detection. 
We have observed that detection by MSA had a similar behavior, except for some channels with $0.05< \delta_\infty < 0.3$, for which MSA is significantly superior to LP. In other words, the transition from LP-improper to LP-proper behavior starts from smaller values of $\delta_\infty$ for MSA.

As an estimate of the probability density function of $\delta_\infty$ for this random construction of the channel response, its histogram has been included in the lower plot of Fig. \ref{d_inf}.

\section{Combined Equalization and LDPC Decoding}
One of the main advantages of the graph-based detection proposed in Section II is that it lends itself to the combining of the equalization with the decoding of an underlying error-correcting code. In this section, we study this joint detection scheme using both the linear programming and the MP algorithms.

\subsection{Coded Linear Programming Detection}
Joint LP equalization and decoding, i.e., the linear relaxation of (\ref{ILP}), is a linear optimization problem in the form of the uncoded LP detection, with the addition of the linear inequalities corresponding to the relaxation of the parity-check constraints of the code. These new constraints cut off from the feasible polytope some of the fractional vertices that can trap the uncoded detection problem, but they also add new fractional vertices to the polytope.
It is not easy to derive general conditions for the success or failure of this problem.
However, we can make the following generalization of Theorem \ref{infinite improper}:
\begin{corollary}
\label{coded LP}
Consider a linear code with no ``trivial'' (i.e., degree-1) parity check, used on a channel satisfying $\delta_\infty<0$, where $\delta_\infty$ is defined in (\ref{def d_inf}). Then, coded LP detection on this system has a non-diminishing WER for large block lengths.
\end{corollary}
\begin{proof}
We have shown in Section III-C that if this condition is satisfied, the all-$\frac{1}{2}$ vector will have a lower cost than the transmitted vector with high probability. 
\footnote{The derivation of the limit of $\delta$ was based on the assumption that the transmitted sequence is an i.i.d. sequence, so that (\ref{bound rho}) holds. While the transmitted sequence is no longer i.i.d. in the presence of coding, we implicitly assume that (\ref{bound rho}) still holds. This is a sufficiently accurate assumption for all codes of practical interest. In particular, (\ref{bound rho}) can be proved for a random ensemble of LDPC codes.} 
It is now enough to show that the all-$\frac{1}{2}$ vector will not be cut off by any error correcting code. To see this, consider a relaxed parity-check inequality of the form 
\begin{equation}
\label{constraints2}
\sum_{i\in V} x_i -\sum_{i\in N_c\backslash V} x_i \leq |V|-1,\ \ \forall\ V\subset N_c\ \text{s.t.}\ |V|\ \text{is odd},
\end{equation}
where $N_c\geq 2$.
To prove that this constraint is satisfied by the all-$\frac{1}{2}$ vector, we consider two cases: $|V|=1$, and $|V|\geq 2$.
If $|V|=1$, the first sum in (\ref{constraints2}) will be equal to $1/2$, and the second sum will be greater than or equal to $1/2$, since $N_c\backslash V$ has at least one member. Hence, the left-hand side of (\ref{constraints2}) will be less than or equal to $|V|-1=0$. Also, if $|V|\geq 2$, the first sum will be equal to $|V|/2\leq |V|-1$, while the second sum is non-negative. Therefore, the left-hand side of the inequality will be less than or equal to its right-hand side. Consequently, in both cases (\ref{constraints2}) will be satisfied. 
\end{proof}

\subsection{Coded Message-Passing Detection}
Similar to LP detection, MP detection can be extended to coded systems by adding the parity-check constraints of the code to the PR Tanner graph, as shown in Fig. \ref{PR-LDPC}, and treating it as a single Tanner graph defining a linear code. 
Despite many similarities, LP and MP decoding schemes have a different nature, which makes them behave differently when used for joint equalization and detection. For example, we cannot derive a conclusion similar to Corollary \ref{coded LP} for MP detection. On the contrary, we have observed in the simulation results that there are LP-improper channels for which coded MP detection does not inherit the undesirable performance of uncoded MP detection.

In this work, we use both the min-sum algorithm and the sum-product algorithm (SPA) for the implementation of coded MP detection. Similar to the uncoded case, as the objective coefficients of MSA, we use the same coefficients as those of LP detection, i.e., $\{q_t\}$ and $\{\lambda_{t,j}\}$, since MSA is invariant under the scaling of the coefficients. For SPA, we observe that each $q_t$ contains a Gaussian noise term with variance proportional to $\sigma^2$. Hence, one can argue that a suitable normalization of the objective coefficients to estimate the ``equivalent LLRs'' is to multiply all the objective coefficients by $2/\sigma^2$. An advantage of this normalization is that, in the absence of ISI, the normalized coefficients become the true LLR of the received samples. In this work, we have used the equivalent LLRs obtained by this normalization for SPA detection.

MSA and SPA decoding are, respectively, approximations of ML and a posteriori probability (APP) detection on the Tanner graph defining the code. These approximations becomes exact if the messages incoming to any node are statistically independent. This happens if the Tanner graph is cycle-free and the channel observations (i.e., the a priori LLRs) are independent. In the proposed graph-based detection neither of these two conditions is satisfied. In particular, the PR layer of the graph contains many 6-cycles, and the channel observations, $\{q_t\}$, are the results of matched filtering of the received signal, and thus contain colored noise. 

\begin{figure}
\centering
\includegraphics[width=3.0 in] {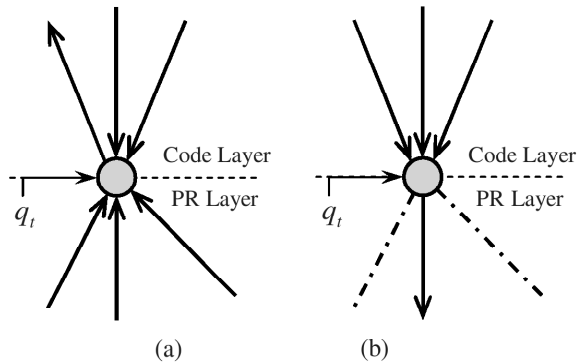}
\caption{Selective message passing at the information bit nodes: (a) Calculating a message outgoing to the code layer (b) Calculating a message outgoing to the PR layer.} \!\!\!\!\!\!\!\!\!
\label{selective MP}
\end{figure}

In order to mitigate the positive feedback due to the cycles of the PR layer, we propose \emph{selective message passing} for coded detection, which has a modified combining rule for messages at the information bit nodes. This modified combining scheme is illustrated in Fig. \ref{selective MP}. In this method, the message outgoing from an information bit node through an edge $e$ in the code layer is computed as a combination of the channel observation and the messages incoming to the bit node through all edges, except edge $e$. On the other hand, the message outgoing through an edge in the PR layer is a combination of the channel observation and messages incoming through only the edges in the code layer. Since there are no 4-cycles in the PR layer of the graph, this modification blocks any closed-loop circulation of messages inside the PR layer. In other words, message passing inside the PR layer will become loop-free. However, there still remain cycles in the code layer, as well as cycles that are generated by combining the code and PR layers. In our simulations, selective MP performed as an effective tool for improving coded MP detection of some channels with undesirable properties, such as the EPR4 channel.

\subsection{Simulation Results}
In this subsection, we present simulation results of coded detection in the presence of ISI using the schemes proposed in this section. In all cases, we have used a regular LDPC code of length 200, rate $1/4$, and variable degree 3. The following PR channels have been used in these simulations: 
\begin{enumerate}
\item \textbf{No-ISI Channel:} $h(D)=1$,
\item \textbf{EPR4 Channel:} $h(D)=1+D-D^2-D^3$ ($\delta_\infty=0$, LP-improper),
\item \textbf{Modified EPR4:} $h(D)=1+D-D^2+D^3$ ($\delta_\infty=\frac{1}{2}$, asymptotically LP-proper),
\item \textbf{PR4 Channel:} $h(D)=1-D^2$ ($\delta_\infty=\frac{1}{2}$, LP-proper).
\end{enumerate}

\begin{figure}
\centering
\includegraphics[width=4.0 in] {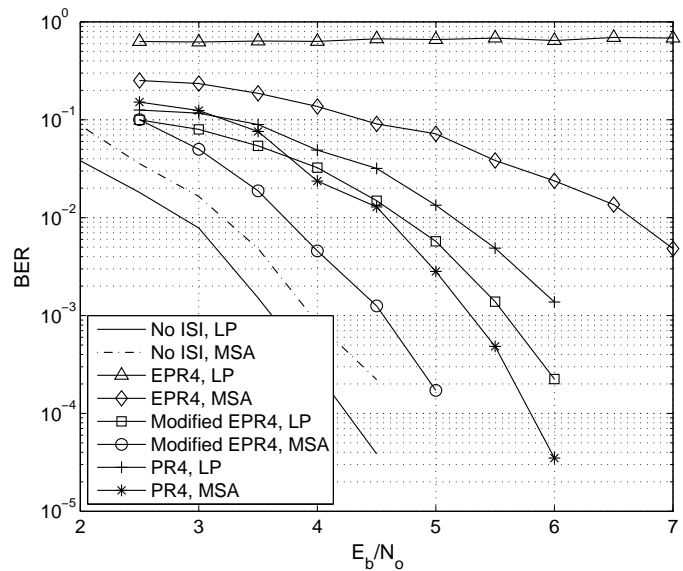}
\caption{BER vs. $E_b/N_o$ for coded LP detection and coded MSA detection in four channels.} \!\!\!\!\!\!\!\!\!
\label{JointDetection}
\end{figure}

In Fig. \ref{JointDetection}, the BER of coded LP and MSA detection has been plotted versus $E_b/N_o$ for the above channels. For all channels, except in the ISI-free case, coded MSA detection outperforms coded LP detection. In particular, for the EPR4, coded LP detection has a BER of about $1/2$ for all SNR values, as predicted by Corollary \ref{coded LP}, while coded MSA detection has a monotonically-decreasing BER.

\begin{figure}
\centering
\includegraphics[width=4.0 in] {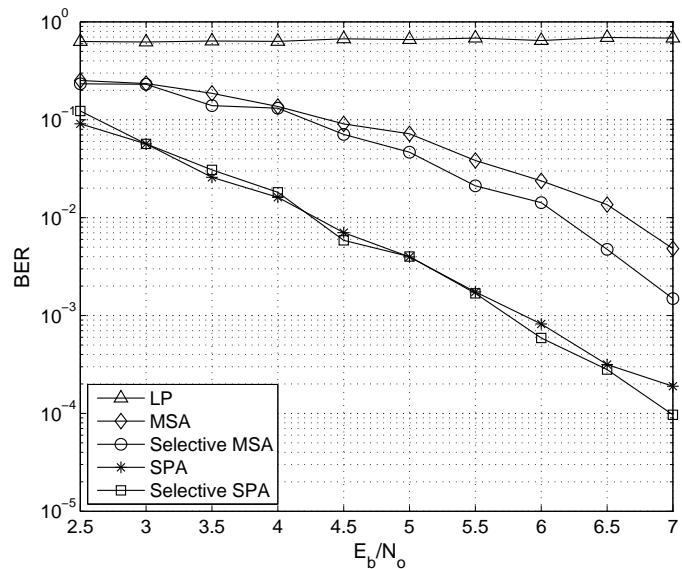}
\caption{BER vs. $E_b/N_o$ for various coded detection schemes in the EPR4 channel.} \!\!\!\!\!\!\!\!\!
\label{JointDetectionEPR4}
\end{figure}

To study the behavior of the different detection methods for the EPR4 channel in more detail, in Fig. \ref{JointDetectionEPR4}, the BER has been plotted versus $E_b/N_o$ for LP, MSA, selective MSA, SPA, and selective SPA. One can observe in this figure that selective message passing is mostly effective for MSA, for which there is a 0.5 dB SNR gain. In addition, by using SPA instead of MSA, we obtain a 2 dB SNR gain. We have observed that, for the other three channel, the gap between MSA and SPA was between 0.3 to 0.7 dB.

% \section{Applications}
\section{Conclusion}
In this paper, we introduced a new graph representation of ML detection in ISI channels, which can be used for combined equalization and decoding using LP relaxation or iterative message-passing methods. By a geometric study of the problem, we derived a necessary and sufficient condition for the equalization problem to give the ML solution for all transmitted sequences and all noise configurations under LP relaxation. Moreover, for certain other channels violating this condition, the performance of LP is very close to that of ML at high SNRs. For a third class of channels, LP detection has a probability of failure bounded away from zero at all SNR, even in the absence of noise. In a step toward the analysis of the performance in the general case, we derived a distance, $\delta_\infty$, for ISI channels, which can be used as a tool to estimate the asymptotic behavior of the proposed detection method. Simulation results show that message-passing techniques have a similar performance to that of LP for most channels. In addition, we studied graph-based joint detection and decoding of channels with LDPC-coded inputs. Simulation results indicate that, in contrast to the uncoded case, message-passing detection significantly outperforms LP detection for some channels.
%Some potential future works include performance analysis of combine equalization and decoding under message-passing using density evolution techniques, and studying the applications of the proposed method for PRML detection and 2D ISI channels.

\end{document}